# Thin chalcogenide capillaries as efficient waveguides in the mid-IR - THz spectral ranges

**Anna Mazhorova, Andrey Markov, Bora Ung, Mathieu Rozé, Stepan Gorgutsa, and Maksim Skorobogatiy[*]**

*École Polytechnique de Montréal, Génie Physique, Québec, Canada*
[*]*maksim.skorobogatiy@polymtl.ca*

**Abstract:** We present chalcogenide glass $As_2Se_3$ capillaries as efficient waveguides in the mid-IR and THz spectral ranges. The capillaries are fabricated using a double crucible glass drawing technique. The wall thickness of the glass capillary is properly designed and controlled during drawing, and we are able to produce capillaries with different wall thickness, starting from 12 µm and up to 130 µm. Such capillaries show low loss properties in the whole target wavelength region. In the mid-IR range guidance is governed by Fresnel reflection and antiguidance mechanisms (ARROWs), while in the THz spectral range thin walls capillaries guide via total internal reflection.

**OCIS codes:** (060.2390) Fiber Optics; infrared (230.7370) Waveguides, (300.6495) Spectroscopy, terahertz

---

## References and links

## 1. Introduction

Chalcogenide glasses are materials of high interest for optical applications in a broad spectral range (1-14 μm) [1], where silica is not fully transparent. The fiber geometry opens new possibilities for a large number of applications not only in mid-IR spectral range, like optical sensing [2] and single-mode propagation of IR light, but also in CO and CO2 laser power delivery systems[3].

In this paper, we propose a chalcogenide glass $As_2Se_3$ fiber in the form of a capillary for the mid-infrared and THz region which are useful in the variety of applications. The capillaries are drawn from a glass preform by using a double crucible glass drawing technique [4,5], and thus, a long fiber is easily produced. The wall thickness of the glass capillary is properly designed and controlled during drawing and we are able to produce capillaries with different wall thickness, starting from 130 μm (Fig.1 (a)) and up to 12 μm (Fig.1(b)).

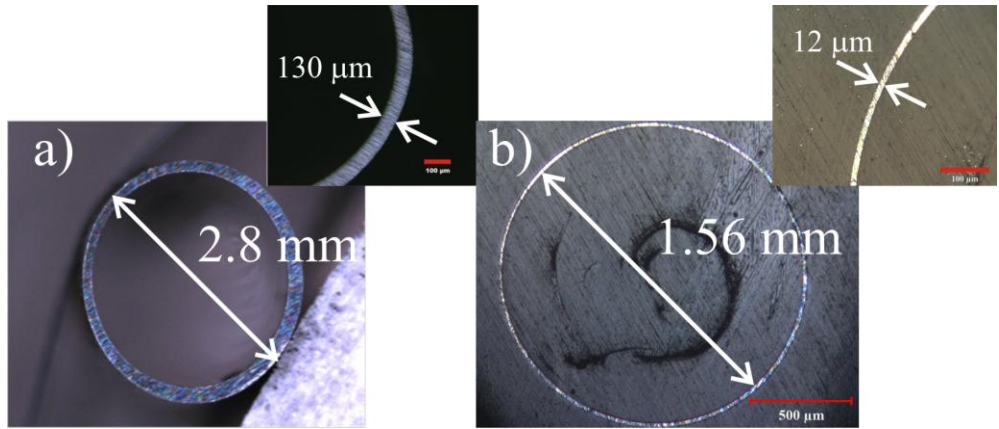

Fig. 1. Fabricated chalcogenide glass capillaries with different wall thickness, starting from (a) 130 μm and up to (b) 12 μm by double crucible glass drawing technique. To perform an imaging of thin wall capillaries, such as capillary with 12 μm walls shown at (b), they were glued in epoxy and then polished.

Guiding mechanism of such capillaries is well known [6-8]. In the THz spectral range they guide via total internal reflection, while in the mid-IR spectral range guiding mechanism is driven by an antiresonant reflecting [9,10]. Wall of the capillary acts as a Fabry–Perot resonator; at the resonant frequencies, the field crosses the material interface, while it is strongly reflected under the antiresonant condition leading to field confinement into the hollow core. Transmission windows are delimited by the resonant frequencies that can be accurately predicted.

## 2. Glass synthesis and capillaries fabrication

Chalcogenide glasses are synthesized usually from single chemical elements, while bulk samples of vitreous arsenic chalcogenides of optical grade are produced by solidification of the glass forming melt [11]. We use $As_{38}Se_{62}$ base glass composition. This composition was chosen because of its higher stability against crystallization during the drawing compared to $As_2Se_3$ ($As_{38}Se_{62}$). Preform of 80 g of the glass was prepared from the pure materials (Se: 99.999% 5N and As: 99.999% 5N) by introducing them in a chemically cleaned and dried silica ampoule. After that the ampoule was pumped using a diffusion vacuum pump to the pressure of $5 \cdot 10^{-7}$ mbar and placed into a rocking furnace where the glass is melted at the temperature above 600°C. At the next step glass was, first, slowly cooled down to 450°C and then quenched in cold water in order to solidify the glass forming-melt. The obtained glass rod is then annealed under the temperature close to glass transition temperature $T_g$ ($T_g =$ 165°C) overnight to reduce inner stress caused by the quenching. After the quenching chalcogenide glass rod typically has smaller diameter than the silica ampoule, thus it might be carefully removed from the ampoule and introduced inside the silica double crucible. Up to now, double crucible set-up was mainly used to draw step index fibers [11]. Both volumes of the double crucible usually contain a core and clad glasses with different refraction index in order to get multimode or single-mode fiber, depending on the core diameter in the final fiber.

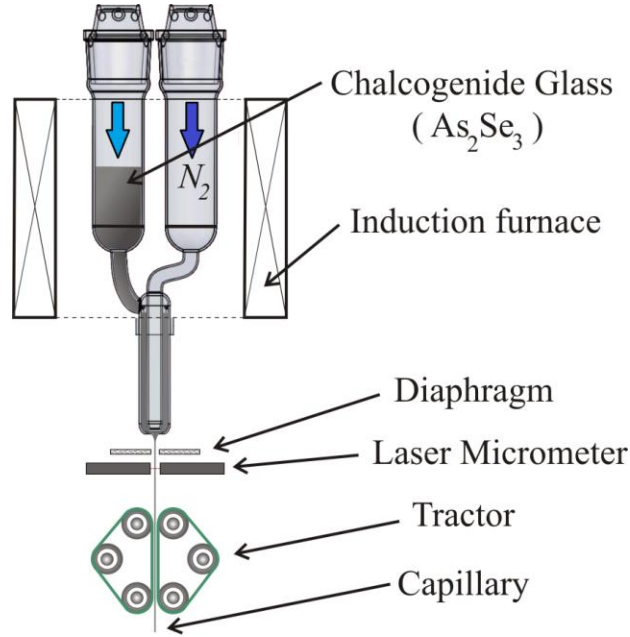

Fig. 2. Set-up for drawing capillaries using the double-crucible method.

Schematic of the typical drawing tower for the capillaries fabricated using the double-crucible method is presented in Figure 2. In our case, the volume of the crucible designed for the fiber core remains empty while the clad crucible contains the synthesized $As_{38}Se_{62}$ glass. Beforehand, the entire chamber with the double crucible is purged at 50°C overnight with a constant flow of $N_2$ to avoid $O_2$ and particles in the heated zones. The temperature is slowly increased from the bottom to the top of the crucible. In the higher zone, at 400°C the glass starts to melt and flows down to the cooler zones (~295°C) along the crucible, thus the viscosity of the melt in the lower zones is higher which in order to have a slow and controllable flow. At the tip of the double crucible a drop of glass is formed and as it goes down by its own weight, the fiber starts to be drawn like in conventional fiber drawing techniques. As soon as the drop reaches the tractor it is pulled down with a controlled speed; meanwhile the pressure in the core crucible is increased in order to form a capillary. By varying the drawing conditions such as tractor speed and pressure in the core, numerous dimensions of capillaries and wall thickness have been obtained.

### 3. Absorption losses of the $As_2Se_3$ glass

Key requirement for the manufacturing transparent optical glass fiber is a highly purified from gas-forming impurities (oxides, OH– group, water, CO, CO2, SeH) initial glass perform as these impurities have absorption bands in the middle and far IR ranges [11]. Fig. 3 presents absorption losses of the $As_2Se_3$ rod with diameter 350 μm and 63 cm length in the 3–5 μm spectral range. Absorption bands in this spectral range (impurity absorptions due to SeH bond at 3.53, 4.12, 4.57 μm) arise due to hydrogen impurities, that are still present after even drastic purification of the glass by multiple distillations [12]. Increase of the optical losses in the 10–12 μm spectral region is usually explained by the intrinsic multi-phonon absorption attributed to the As–Se and As–Se–As bonds.

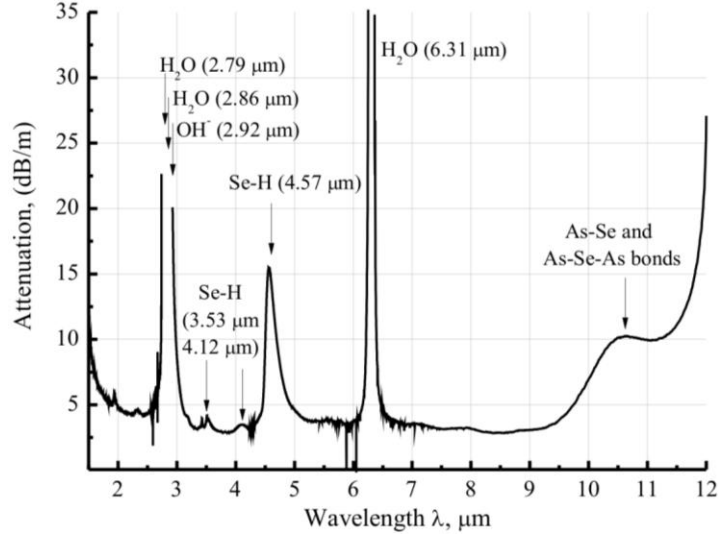

Fig. 3 Absorption losses of the $As_2Se_3$ rod with diameter 350 μm and 63 cm in length. Absorption bands in the 3–5 μm spectral range corresponds impurity absorptions due to SeH bond at 3.53, 4.12, 4.57 μm. Also big absorption bands at 2.7-2.93 μm and 6.31 μm are due to OH– group and water

## 4. Results and discussion

Guiding mechanism of the leaky core modes in the capillary waveguide (chalcogenide glass tube) is similar to that of the antiresonant reflecting optical waveguides (ARROWs) [13], and can be briefly described by viewing the capillary wall as a Fabry-Perot etalon. At the resonate frequencies (or close to them) nearly no reflection takes place at the interior side of the tube and thus fields could hardly exist inside the capillary. On the other hand, under the antiresonant conditions, i.e. at the frequencies away from the resonant ones, considerable reflections at the interior side of the tube results into the core modes (waves bounce back and forth inside the capillary). The resonant frequencies can be accurately predicted with following relation:

$$f_m = \frac{mc}{2t\sqrt{n^2_{high-index\_wall} - n^2_{low-index\_air}}}. \quad (1)$$

where $c$ is the speed of light in vacuum and $m$ is an integer and $n$ is refractive indices of chalcogenide glass and air. Due to resonances transmission (and absorption losses) through the thin-walled capillaries shows periodic minimums and maximums as a function of frequency. The bandwidth of the transmission windows depends on the chalcogenide glass refractive index and thickness of the wall:

$$\Delta f = f_{m+1} - f_m = \frac{c}{2t\sqrt{n^2_{high-index\_wall} - n^2_{low-index\_air}}}. \quad (2)$$

In wavelength:

$$\Delta \lambda = \lambda_{m+1} - \lambda_m = \frac{\lambda^2}{2t\sqrt{n^2_{high-index\_wall} - n^2_{low-index\_air}}}. \quad (3)$$

Modal characteristics of the pipe waveguides have been investigated using exact transfer matrix theory and finite element method in Comsol software. For each capillary the attenuation constants of the first lowest modes of the waveguide were calculated. The main characteristics of the higher order modes are the same as those of the fundamental $HE_{11}$ mode including the same resonant regions with high propagation losses. Modal indices of the higher order modes are lower than that of the fundamental mode. This means that these modes propagate less parallel than the fundamental one and the smaller value of the incident angle on the core-cladding interface results in higher attenuation constants for these modes. Also we should take into consideration that not all this modes can be excited simultaneously. Coupling of the initial beam into these modes is defined by its state of polarization. For example, for the linearly polarized beam coupling to TE and TM modes is equal to zero.

*4.1 Capillaries in Mid-IR spectral range*

Transmission measurements were preformed in two spectral ranges: in mid- IR from 1.5 to 14 μm by Fourier Transform Infrared Spectrometry (FT-IR spectrometer) and in THz sprectral range from 0.1 up to 3 THz by modified THz-TDS (Time-Domain Spectroscopy) setup. The setup consists of a frequency-doubled femtosecond fiber laser (MenloSystems C-fiber laser) used as a pump source and identical GaAs dipole antennae used as source and detector yielding a spectrum ranging from ~0.1 to 3 THz. Contrary to the standard THz-TDS setup where the configuration of parabolic mirrors is static, our setup has mirrors mounted on translation rails. This flexible geometry facilitates mirrors placement, allowing measurement of waveguides up to 50 cm in length without realigning the setup.

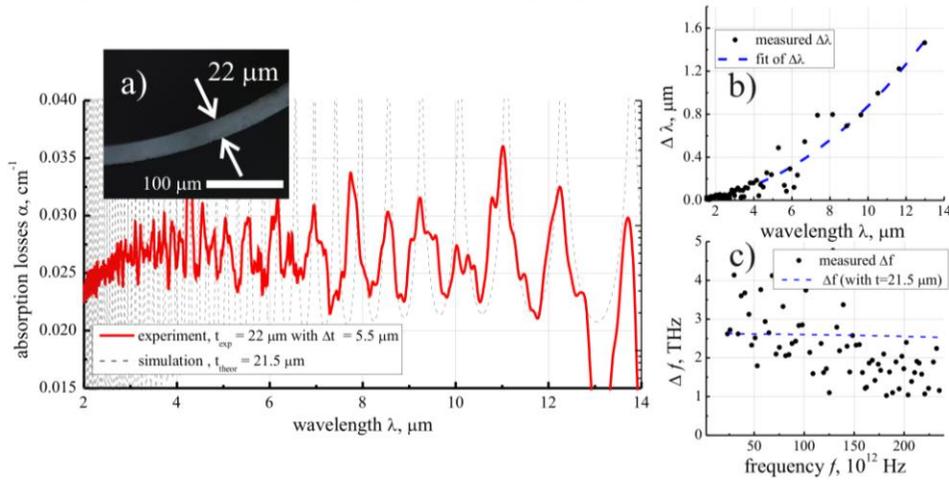

Fig. 4 (a) absorption losses of capillary with averaged wall thickness 22 μm in spectral range from 1.5-14 μm, measured with cut-back method. (b) measured period of the resonances $\Delta\lambda$ as a function of wavelength $\lambda$ (μm) from Fig. 4 a) of absorption losses. Blue dashed line is the experimental fit of the resonances, which gives the wall thickness value of $t = 21.2 \pm 1.3$ μm, which is in a good agreement with measured with optical microscope wall thickness $t = 22.1 \pm 5.5$ μm.

In mid-IR capillary guidance mechanism is anti-resonant reflection from the capillary walls with most power concentrated inside of the capillary hollow core. Very thin chalcogenide capillaries with wall thickness below 20μm and outer diameter 1-2 mm guide very well mid-IR radiation with losses in the 7-16 dB/m (0.025 cm$^{-1}$) depending on the frequency of observation. Fig. 4 a) illustrates absorption losses of capillary with averaged wall thickness 22 μm in spectral range from 1.5-14 μm, measured with cut-back method. The result of the exact transfer matrix calculation of the fundamental mode is shown as a dashed line. The obtained result shows that for such long capillaries the propagation can be considered single-mode. Fig. 4 b) shows measured period of the resonances $\Delta\lambda$ as a function of wavelength $\lambda$ (μm) from Fig. 4 a) of absorption losses. Blue dashed line is the experimental fit of the resonances, which gives the wall thickness value of $t_{fit} = 21.2 \pm 1.3$ μm, which is in a good agreement with measured with optical microscope wall thickness $t = 22.1 \pm 5.5$ μm. Compared to the thin capillaries, spectral oscillations of the capillary with twice thick walls are much less pronounced especially at shorter wavelengths. Weak oscillations are still visible between 8 and 13μm. Fig. 5 a) illustrates absorption losses of capillary with averaged wall thickness 40 μm in spectral range from 1.5-14 μm. Fig. 5 b) shows measured period of the resonances $\Delta\lambda$ as a function of wavelength $\lambda$ (μm) from Fig. 5 a) of absorption losses, blue dashed line is the experimental fit of the resonances, which gives the wall thickness value of $t_{fit} = 41.4 \pm 1.8$ μm, which is in a good agreement with measured with optical microscope wall thickness $t = 40.1 \pm 6.2$ μm.

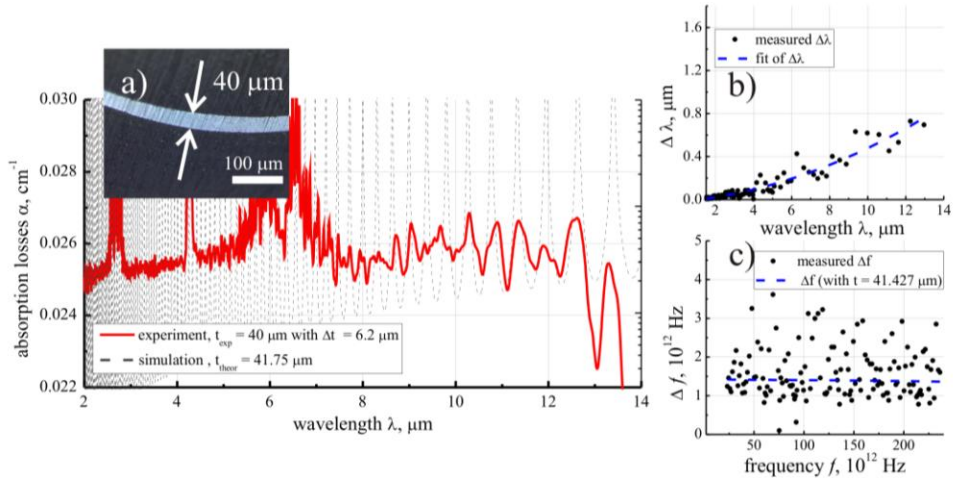

Fig. 5. (a) Absorption losses of capillary with averaged wall thickness 40 μm in spectral range from 1.5-14 μm. (b) measured period of the resonances $\Delta\lambda$ as a function of wavelength $\lambda$ (μm) from Fig. 5 a) of absorption losses. Blue line is the experimental fit of the resonances, which gives the wall thickness value of $t = 41.4 \pm 1.8$ μm, which is in a good agreement with measured with optical microscope wall thickness $t = 40.1 \pm 6.2$ μm.

When using thicker capillaries, contrast between minimums and maximums in the mid-IR transmission spectrum decreases and in very thick capillaries spectral with wall thickness above 100 μm oscillations are not detectable with FTIR technique. Capillaries tend to show a featureless spectrum with ~17 dB/m (0.037 cm-1) losses.

*4.2 Terahertz spectral range*

In the THz spectral range the thicker capillaries (~100 μm) show clear anti-resonant guidance with periodic minimums and maximums as a function of frequency. Figure 6 (a) illustrates the transmittance through 50 cm long capillary with wall thickness 98 μm. The first peak of transmittance at the frequency 0.26 THz corresponds to the propagation of the total internal reflection mode, at such frequency the conditions for the exciting of the ARROW mode are not met, since the wavelength is comparable with the diameter of the pipe. All other peaks correspond to the ARROW propagation regime. The numerical simulation shows that the capillary of 50 cm length operates in almost single-mode (HE11) regime, higher-order modes contribute only in small oscillations of the transmittance curve due to their higher attenuation constants. Thinner capillaries (<40um) show a very broad and featureless transmission In the THz spectral range. Losses are low (below 0.2cm-1) and guidance mechanism is a total internal reflection. The capillaries guide as a subwavelength fiber as capillary thickness is much smaller than the wavelength of light.

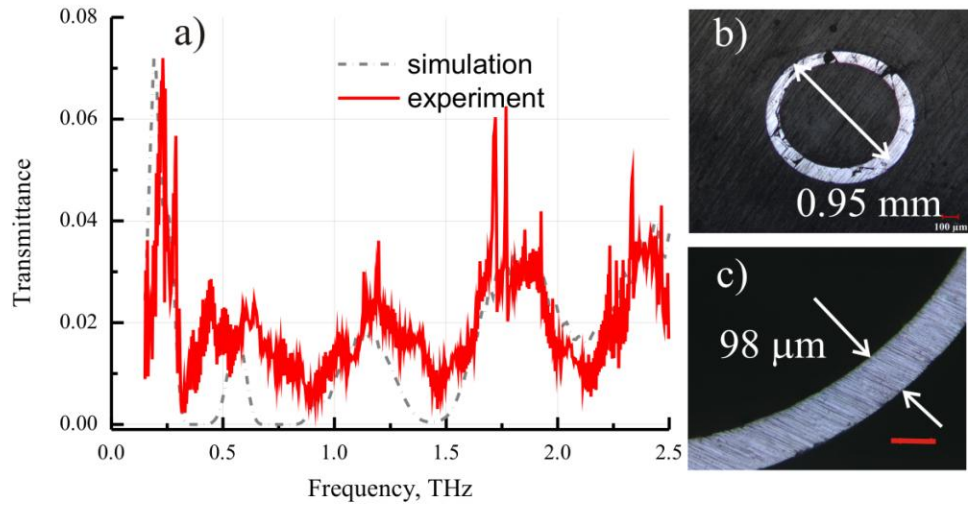

Fig. 5. (a) Absorption losses of capillary with averaged wall thickness 133 µm in spectral range from 0-2.5 THz b,c) capillary used in the experiments with outer diameter of 2.8 mm and averaged wall thickness 133 µm .

**Conclusion**

We present chalcogenide glass $As_2Se_3$ capillaries as efficient waveguides from THz to mid-IR range and that are fabricated by using a double crucible glass drawing technique. The wall thickness of the glass capillary is properly designed and controlled during drawing, and we are able to produce capillaries with different wall thickness, starting from 130 µm and up to 12 µm. Such capillaries show low loss properties in the whole target wavelength region. However, in the mid-IR range guidance is governed by Fresnel reflection and antiguidance mechanisms (ARROWs), while in the THz spectral range thin walls capillaries guide via total internal reflection.